\documentstyle[aps,epsf,preprint,graphicx,prl]{revtex}

\begin{document} 
\draft

\def\beq{\begin{equation}}
\def\eeq{\end{equation}}
\def\beqn{\begin{eqnarray}}
\def\eeqn{\end{eqnarray}}
\def\btimes {\mbox{\boldmath $\times$}}
\def\bbox {\mbox{\boldmath $\box$}}
\def\ed{\end{document}}

\def\veps {{\varepsilon}}
\def\I {{\bf I}}
\def\II {{\bf II}}
\def\III {{\bf III}}
\def\IV {{\bf IV}}
\def\V {{\bf V}}
\def\VI {{\bf VI}}
\def\J {{\bf J}}
\def\H {{\bf H}}
\def\E {{\bf E}}
\def\1 {{\bf 1}}
\def\2 {{\bf 2}}
\def\3 {{\bf 3}}
\def\r {{\bf r}}
\def\k {{\bf k}}
\def\p {{\bf p}}
\def\n {{\bf n}}
\def\A {{\bf A}}
\def\bv {{\bf v}} 
\def\AAN {$\!\!\!$ A$^{^{\!\!\!\!\! {\tiny {\circ}}}}$}
\def\aaN {$\!\!$ a$^{^{\!\!\!\! {\tiny {\circ}}}}$}

\title{Continuous configuration-interaction for condensates in a ring}

\author{Ofir E. Alon\footnote{E-mail: ofir@tc.pci.uni-heidelberg.de}, Alexej I. Streltsov, Kaspar Sakmann,
and Lorenz S. Cederbaum}
\address{Theoretische Chemie, Physikalisch-Chemisches Institut, Universit\"at Heidelberg,
Im Neuenheimer Feld 229, D-69120 Heidelberg, Germany}

\maketitle

\begin{abstract}
A continuous configuration-interaction approach for condensates in a ring is introduced.
In its simplest form this approach utilizes for attractive condensates 
the Gross-Pitaevskii symmetry-broken solution and arrives at a ground-state of correct symmetry.
Furthermore, the energy found is {\it lower} than the Gross-Pitaevskii one
and, with increasing number of particles and/or strength of inter-particle interaction,
is even {\it lower} than that accessed by tractable diagonalization of the many-body Hamiltonian.
Implications also to excited states are briefly discussed.
\end{abstract}

The physics of bosons in low dimensions has regained much attention recently  
following the realization of Bose-Einstein condensates in effectively lower-dimension trap potentials \cite{Gorlitz} 
and observation of bright solitons for attractive condensates in a quasi-one-dimensional (1D) trap \cite{Strecker} and 
optical waveguide \cite{Khaykovich}.
Attractive condensates in 1D do not undergo collapse and have been 
a subject of long interest, see, e.g., 
Refs.~\cite{LL_appen,Calogero,Muga,Ueda_PRL1,Reinhardt1,Kavoulakis,Ueda_PRA1}.
For a system of two \cite{LL_appen} and three \cite{Muga} attractive bosons in a ring,
although it is quite involved, 
it is possible to obtain expressions for the exact ground-state energy and wavefunction.
However, for more than a few bosons and as the inter-particle interaction is increased,
it becomes impossible to compute the {\it exact} ground-state energy and many-body wavefunction,
and thus approximations are a must.

The Gross-Pitaevskii (GP) equation has been a very successful approximation for Bose-Einstein condensates 
and can explain many experiments with them, see, e.g., Refs.~\cite{review1,review2} and references therein.
For attractive condensates in a ring, 
due to the continuous axial symmetry,
the GP equation has been solved analytically \cite{Reinhardt1}.
Here, for weakly-interacting bosons the ground-state solution is symmetry-preserving,
i.e., it possesses the rotational symmetry of the {\it full} many-body problem.
For stronger (attractive) interactions, the GP solution lowest in energy becomes symmetry-broken
and the gain in energy in comparison to the symmetry-preserving solution is {\it substantial}, see inset of Fig.~1. 
Because the GP orbital breaks the symmetry of the many-body system,
it cannot properly describe the true ground-state. 

In this Letter, a continuous configuration-interaction (CI) approach for condensates in a ring is introduced.
For the attractive condensate, we will start from the GP symmetry-broken solution and  
obtain a many-body wavefunction of correct symmetry and energy {\it lower} than the GP (symmetry-breaking) one.
In addition as we explain and demonstrate below,
with increasing number of particles and/or inter-particle interaction,
our simple approach leads to an energy which is even {\it lower} than that accessed by tractable 
diagonalization of the many-body Hamiltonian.
While solving explicitly for the ground state of attractive condensates in a ring,
implication of our approach also to excited states
and in higher dimensions will be discussed in brief. 

Our starting point is the dimensionless many-body Hamiltonian describing $N$ interacting bosons in a ring \cite{Ueda_PRA1}:
\beq\label{Ham_MB}
 \hat H = \sum_i \hat L_{\varphi_i}^2  + \frac{U}{2} \sum_{i\ne j} \delta(\varphi_i-\varphi_j), \ \
 \hat L_\varphi = -i \frac{\partial}{\partial \varphi}
\eeq
Here, $\varphi$ is the azimuthal angle and $U$ describes the effective boson--boson interaction.
It is convenient to introduce the dimensionless parameter $\gamma = \frac{U(N-1)}{2\pi}$ which is negative for attractive condensates. 

The simplest approximation to the ground-state of Hamiltonian (\ref{Ham_MB})
is the Hartree approximation where the many-body wavefunction is given by $\Phi(\vec \varphi) = \Pi_i \, \phi(\varphi_i)$.
Minimizing the energy for this wavefunction leads to the GP orbital $\phi(\varphi)$ \cite{review1,review2}.
To proceed, we briefly examine the GP solution $\phi(\varphi)$ 
(with periodic boundary conditions $\phi(0)=\phi(2\pi)$) \cite{Reinhardt1,Ueda_PRA1}.
Below the critical value $|\gamma|\le 0.5$, $\phi(\varphi)=1/\sqrt{2\pi}$, i.e., it is a constant.
Consequently, the Hartree wavefunction $\Phi(\vec \varphi)$
possesses the symmetry of the exact ground-state of (\ref{Ham_MB}),
i.e., it is an eigenfunction of the total angular-momentum operator,
$\hat L_{\vec \varphi} = \sum_i \hat L_{\varphi_i}$, with the eigenfunction $0$.
For $|\gamma|>0.5$, as the number of particles and/or effective interaction is increased,
$\phi(\varphi)$ is given by a Jacobian elliptic function \cite{Reinhardt1,Ueda_PRA1},
namely, it starts to localize and becomes a symmetry-broken solution.
In other words, to gain energy with respect to symmetry-preserving solution in this parameter regime 
the mean-field treatment ``prefers'' to break the rotational symmetry of the many-body Hamiltonian (\ref{Ham_MB}), see inset of Fig.~1.
Consequently, $\Phi(\vec \varphi)$ is no longer an eigenfunction of $\hat L_{\vec \varphi}$ and thus does not represent
an appropriate approximation for the exact ground state.

The question we would like to address here is how to go {\it beyond}
the Hartree approximation for attractive condensates in a ring.
Of course, it would be of great advantage to start where the GP equation makes a success,
i.e., with the symmetry-broken solution $\phi(\varphi)$.
To this end, we employ the Hartree product $\Phi(\vec \varphi)$ as a many-body ``basis function''
and plug it into a multiconfigurational variational ansatz for the many-body Hamiltonian (\ref{Ham_MB}).
Since the ring has a {\it continuous} symmetry, it is reasonable to construct a
{\it continuous} CI ansatz for the many-body wavefunction.
Namely, we define the continuous CI wavefunction as:
\beq\label{CC_sym_def}
 \Psi(\vec \varphi) = {\mathcal N} \int_0^{2\pi}d\theta \, C(\theta) \, \Phi (\vec \varphi-\theta),
\eeq
where $\Phi(\vec \varphi-\theta) = \Pi_i \, \phi(\varphi_i-\theta)$ and  ${\mathcal N}$ is the normalization constant.
$C(\theta)$ is a continuous function which is determined from the requirement that the 
expectation value of the Hamiltonian (\ref{Ham_MB}) with respect to the wavefunction $\Psi(\vec \varphi)$ is minimized.
Having done that,
an equation (condition) for $C(\theta)$ is obtained
admitting the solution
\beq\label{C_solution}
 C(\theta) = e^{+i L \theta}, \ \ L=0,\pm 1, \pm 2,\ldots.
\eeq
It can be verified by substituting Eq.~(\ref{C_solution}) into Eq.~(\ref{CC_sym_def}) 
that $\Psi(\vec \varphi)$ 
is an eigenfunction of $\hat L_{\vec \varphi}$
with the eigenvalue $L$. 
For the ground state, the minimum of the energy is obtained for $L=0$, i.e., $C(\theta) = 1$.
Thus, starting from the Hartree product $\Phi(\vec \varphi)$ we obtain a simple and attractive many-body 
wavefunction for the ground state.
Following the above, it is an eigenfunction of $\hat L_{\vec \varphi}$ with the eigenvalue $0$
and thus possesses the correct symmetry of the exact ground state of a condensate in the ring.

To get a grasp at the many-body nature of $\Psi(\vec \varphi)$, we construct a specific example. 
For this, we expand the GP orbital in plane waves,
$\phi(\varphi) = \frac{1}{\sqrt{2\pi}} \left[ a_0 + \sum_{l>0} a_l (e^{+il\varphi} + e^{-il\varphi}) \right], \ 
a_0^2 + 2\sum_{l>0} a_l^2 = 1$, where the $a_l$'s are real since $\phi(\varphi)$ is an even and real function of $\varphi$ 
\cite{Reinhardt1,Ueda_PRA1}.
Just above the critical value $|\gamma|=0.5$ we can retain the terms with
$a_0$ and $a_1$ because higher Fourier coefficients, $a_{l>1}$, are smaller.
Constructing the corresponding Hartree product, 
one obtains after some algebra the many-body wavefunction for $N$ bosons
\beqn
 \Psi(\vec \varphi) \!&\simeq&\! {\mathcal N} \frac{1}{(2\pi)^{N/2}} \Bigg[a_0^N + a_0^{N-2}a_1^2
  \sum_{i\ne j} \cos(\varphi_i-\varphi_j) + \nonumber \\  
 \!&+&\! a_0^{N-4}a_1^4  \sum_{i\ne j\ne k\ne l} \cos(\varphi_i+\varphi_j-\varphi_k-\varphi_l) + \ldots
 \Bigg], \
\eeqn
where the summation runs over all non-negative powers of the form $a_0^{N-2j}a_1^{2j}$.
It can be seen that indeed all boson coordinates are coupled to one another and that $\Psi(\vec \varphi)$
is an eigenfunction of $\hat L_{\vec \varphi}$ with the angular momentum $0$.

Having at hand a wavefunction, 
one can proceed and calculate all ground-state properties,
beginning with the energy per particle 
$\varepsilon[\phi] = \frac{1}{N} <\Psi(\vec \varphi)|\hat H| \Psi(\vec \varphi)>$.
In writing $\varepsilon[\phi]$ we keep in mind that $\varepsilon$ is
a functional of the GP orbital $\phi(\varphi)$ (which is given here analytically \cite{Reinhardt1,Ueda_PRA1}).
The variational principle used to derive Eq.~(\ref{C_solution}) ensures that $\varepsilon$ is {\it lower} than the GP one,
$\varepsilon_{\mathrm GP}$.
The evaluation of $\varepsilon$ (and of the normalization constant ${\cal N}$)
involves two integrations (that depend on the difference $\theta-\theta'$ only) 
which are all simple to implement, e.g., 
using the discrete variable representation (DVR) method \cite{DVR}.

In Fig.~1 we present the outcome of calculating the energy $\varepsilon$
relative to $\varepsilon_{\mathrm GP}$
for $N=10,25,50,100,300$ and $500$ particles and $|\gamma|$ extending up to 2.2.
The energy gain our continuous CI ansatz achieves is clearly seen.
For $N \gg 1$, 
we see that $\varepsilon$ approaches $\varepsilon_{\mathrm GP}$ from below.
This is in favor of our ansatz since it is known that in this limit
the GP energy approaches the {\it exact} energy from above in 1D \cite{Calogero}. 

Also shown in Fig.~1 (inset) are the results of diagonalizing the many-body Hamiltonian (\ref{Ham_MB})
with $|l|\le 2$ and $|l|\le 3$ single-particle plane waves for $N=25$ bosons and $|\gamma|=1.25$ and $2.0$.
For these parameters, the diagonalization energies are even higher than $\varepsilon_{\mathrm GP}$,
least to say from our obtained $\varepsilon$.
Namely, with almost no effort we obtained a many-body wavefunction of correct symmetry
and lower energy. 
That with increasing number of particles and/or inter-particle interaction,
our energy is even {\it lower} than that accessed by tractable diagonalization of the many-body Hamiltonian
is further explained below.

Next, we chose to study the fragmentation of the ground state
which has been described recently in Refs.~\cite{Ueda_PRL1,Ueda_PRA1}.
To this end, we need to calculate the eigenvalues of the reduced one-particle density $\rho(\varphi,\varphi')$,
which is directly obtained from the many-body wavefunction $\Psi(\vec \varphi)$.
It is straightforward to prove that, 
since the Hamiltonian (\ref{Ham_MB}) is rotationally symmetric, 
the natural orbitals are plane waves. 
In other words, the reduced one-particle density takes on a simple form
$$ 
 \rho(\varphi,\varphi')  = \frac{1}{2\pi} \sum_l p_l\, e^{+i l (\varphi-\varphi')},
\eqno{(5a)}
$$
where the $p_l$'s are the eigenvalues (occupations).
On top of that, in the large particle limit 
we obtain the appealing result that 
$$
p_l \stackrel{N\gg 1}{\longrightarrow } a_l^2.
\eqno{(5b)}
$$
This means that the information on the fragmentation of the {\it many-body} wavefunction is given in the large particle limit
 by the plane-wave decomposition of the {\it mean-field} symmetry-broken orbital.
In Fig.~2, we present the values of the occupation $p_l$ multiplied by the weight $l^2$ (see explanation below)
for $N=100$ particles and for $|\gamma|$ ranging up to 2.2. 
The fragmentation of the ground state is clearly seen and the inclusion of additional waves
as $|\gamma|$ is increased is shown to be easily described by our continuous CI ansatz.
In addition, the large-particle limit given by Eq.~(5b) is already approached for a number of bosons as low as $100$.

The inherent ``difficulty'' of diagonalizing the many-body Hamiltonian 
(\ref{Ham_MB}) for more than a few bosons is also illustrated in Fig.~2.
For this, let us examine the plane-wave decomposition of the GP solution $\phi$ for $|\gamma|>0.5$.
The contribution of each plane wave to the kinetic energy in $\varepsilon$ is given 
by $l^2 a_l^2 \approx l^2 p_l$ (hence, the scaling by $l^2$ in Fig.~2).
This provides a simple  
estimate to the {\it error} introduced into a typical diagonalization calculation 
as a result of truncating the number of plane waves included
(we checked this estimate in several examples and found it to be quite good).
It can be seen in Fig.~2 that the $l^2 p_l$ quickly become non-negligible for $|\gamma|>0.5$. 
Thus, the {\it shape} of $\phi(\varphi)$ provides also a measure for the computational
effort required to converge a diagonalization calculation of the many-body problem (\ref{Ham_MB}).
As an instructive example, 
we estimate for $|\gamma|=2.0$ what is the number of particles
for which {\it more than} $10^9$ configurations are needed to compute the energy within an absolute error of $0.01$
(which is not too high accuracy).
Taking all waves such that the error introduced by the cutoff in $0.01$  
(in this case we take all waves for which $l\le 8$),
we find this number to be $N=25$.
This illustrates that, as the number of particle and/or inter-particle interaction is increased, 
it becomes impractical to even approach converged diagonalization results for already two dozens of bosons!

So far, we used the GP orbital $\phi(\varphi)$ for the Hartree wavefunction $\Phi(\vec \varphi)$
entering the continuous CI ansatz, see Eq.~(\ref{CC_sym_def}). 
By construction, the GP orbital minimizes the Hartree energy $\varepsilon_{\mathrm GP}$.
We may enquire what is the {\it optimal} orbital that minimizes our continuous CI energy?
This orbital is not the GP solution.
It can be shown that by subtracting an infinitesimal constant from the GP orbital 
the energy is further lowered with respect to the above-computed $\varepsilon$. 
To find this optimized orbital,
we have to minimize the energy functional $\varepsilon[\phi]$ with respect to the orbital $\phi$.
The equation obtained is quite involved and its exploitation is out of the scope of the present work.
We have reasons to expect that this optimization will take us close to the exact solution.
For $N\gg 1$ our solution behaves correctly as discussed above.
At the other end of the scale, namely for two bosons,
we can prove that our ansatz with optimized $\phi(\varphi)$ leads to the exact solution.

Here, to get a physical insight into how the optimized orbital may look like, we discuss 
the following result. 
The above-mentioned linear-response analysis
suggests that a slight {\it localization} of the GP orbital leads to further lowering of the continuous CI energy.
To illustrate this finding, we computed $\varepsilon[\phi]$ with our continuous CI approach for 
$N=10,25,50$ and $100$ bosons and $|\gamma|=0.7$,
where the orbital taken as $\phi$ is now the GP solution for slightly different values of $|\gamma|$, see Fig.~3A.
For $|\gamma|$ slightly larger than $0.7$ the 
corresponding GP
orbital is slightly narrower than the orbital at $|\gamma|$=0.7
and the energy has indeed decreased (after the minimum is reached, the energy of course rises again).
The results presented in Fig.~3A also show this finding to be more pronounced for a smaller number of bosons.

Interestingly, the finding that a small localization of the GP orbital leads to a lower energy 
also applies to the regime $|\gamma|\le 0.5$, where the GP solution is symmetry-preserving.
Specifically, we illustrate in Fig.~3B that a small localization of the {\it constant-valued} GP orbital
at, say, $|\gamma|=0.35$ leads to a decrease in the energy.
Here, we calculated the energy $\varepsilon[\phi]$ with slightly localized
GP orbitals taken from the regime just above $|\gamma|=0.5$, see Fig.~3B.

Finally, let us briefly discuss further implications of our continuous CI approach.
First, one need not restrict himself to Hartree wavefunctions in Eq.~(\ref{CC_sym_def});
Other symmetry-broken many-body wavefunctions can be used there as well.
Secondly, the general solution for $C(\theta)$ in Eq.~(\ref{C_solution})
leads to a many-body wavefunction of any chosen angular momentum $L$.
In particular, one can apply our approach to the lowest excited states of
a given angular momentum $L$ and calculate  
the so-called Yarst line of strongly-interacting attractive condensates. 
Thirdly, the method and analysis works of course for any potential with azimuthal symmetry.
In particular, 
for the 2D isotropic harmonic trap potential and attractive bosons 
our method is useful on similar grounds as discussed above \cite{remark}.
Last but not least, our continuous CI ansatz  
can also be used with time-dependent orbitals.
This generalization is suitable to describe the dynamics of a condensate in 
 time-dependent potentials, e.g., when the radius of the ring (trap)
 or the strength of the inter-particle interaction is oscillating.

In conclusion, a continuous CI approach for bosons in a ring has been presented.
In its simplest form for attractive condensates, 
it builds on a single Hartree configuration 
(evaluated with the GP symmetry-broken solution) 
and continuously rotated around the ring.
The result is a many-body wavefunction of correct symmetry and energy {\it lower} than the GP one.
With increasing number of particles and/or inter-particle interaction,
our simple approach leads to a {\it lower} energy than that accessed by tractable diagonalization of the many-body problem.
Our ansatz for the wavefunction in terms of the GP orbital shows that 
the fragmentation of the ground state 
is simply obtained (with increasing number of particles) from the Fourier transform of this orbital.

\acknowledgments

\noindent
We thank Prof. Masahito Ueda for communicating us unpublished results.

%%%%%%%%%%% Fig. 1

\begin{figure}[ht]
\includegraphics[width=10cm,angle=0]{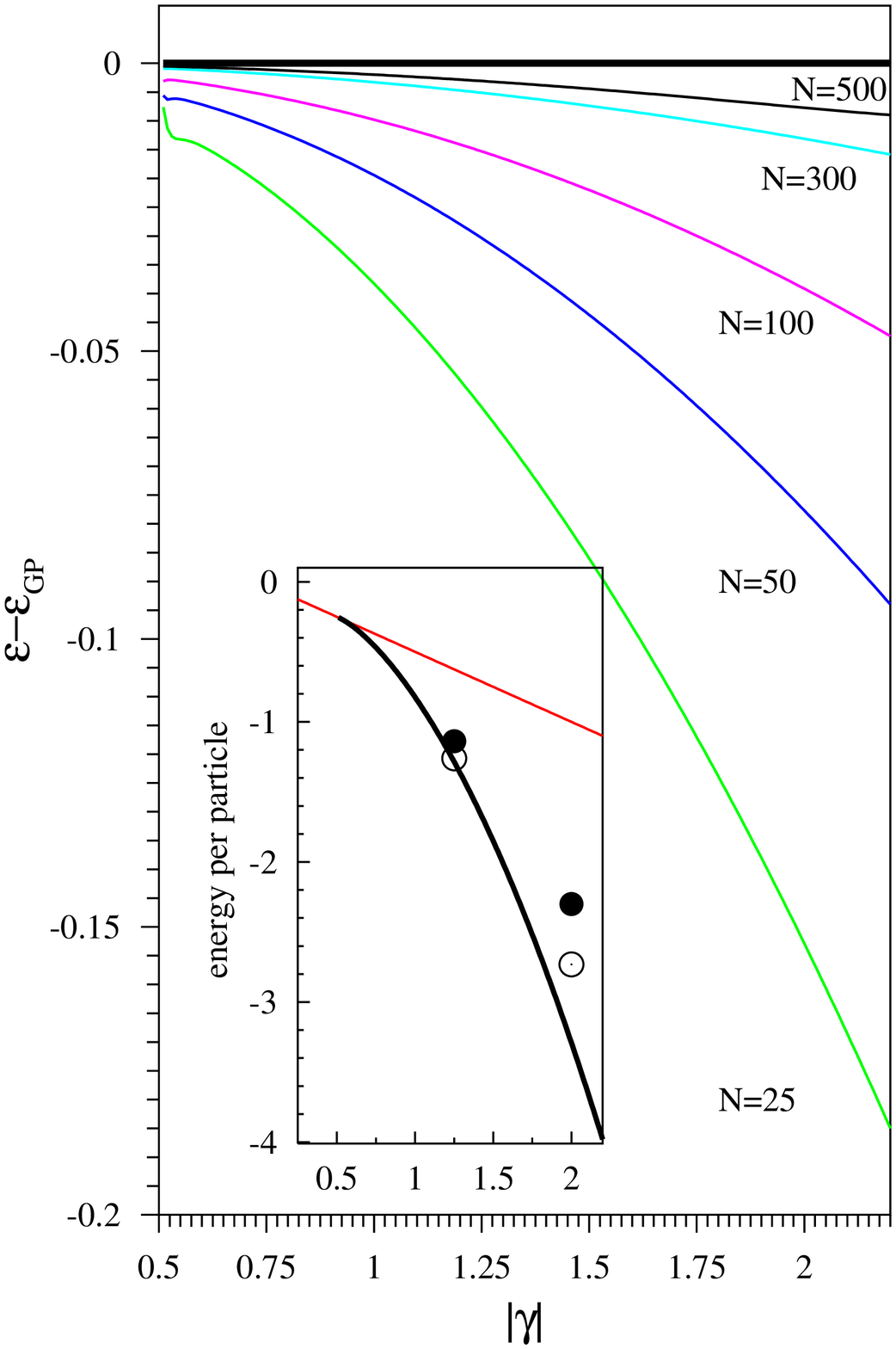}
\caption [kdv]{Ground-state energy per particle of the continuous CI ansatz, $\varepsilon$, 
relative to that of the GP symmetry-broken solution $\varepsilon_{\mathrm GP}$ for $N=10,25,50,100,300$ and $500$ bosons.
The inset shows the energies of the GP symmetry-preserving and symmetry-broken solutions.
For $|\gamma|>0.5$ the solution {\it lowest} in energy becomes symmetry-broken.
Also shown in the inset is the result of diagonalizing the many-body Hamiltonian (\ref{Ham_MB})
with $|l|\le 2$ and $|l|\le 3$ (marked, respectively, by ``$\bullet$'' and ``$\odot$'')
 single-particle plane waves for $N=25$ bosons, and $|\gamma|=1.25$ and $2.0$.
The diagonalization energies are {\it higher} than the GP symmetry-broken ones,
see discussion in text.
}
\end{figure}

\newpage

%%%%%%%%%%% Fig. 2

\begin{figure}[ht]
\includegraphics[width=10cm,angle=-90]{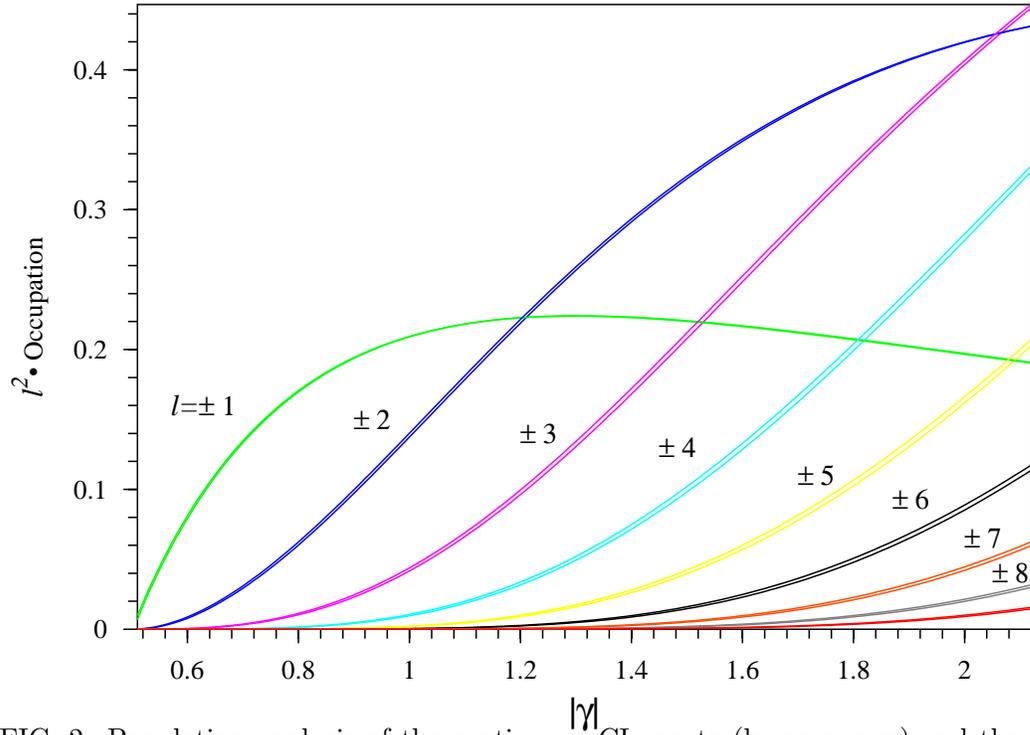}
\caption [kdv]{Population analysis of the continuous CI ansatz (lower curves)
and the Fourier components square of the symmetry-broken GP orbital (upper curves)
for $N=100$ bosons versus $|\gamma|$.
}
\end{figure}

\newpage

%%%%%%%%%%% Fig. 3

\begin{figure}[ht]
\includegraphics[width=10cm,angle=-90]{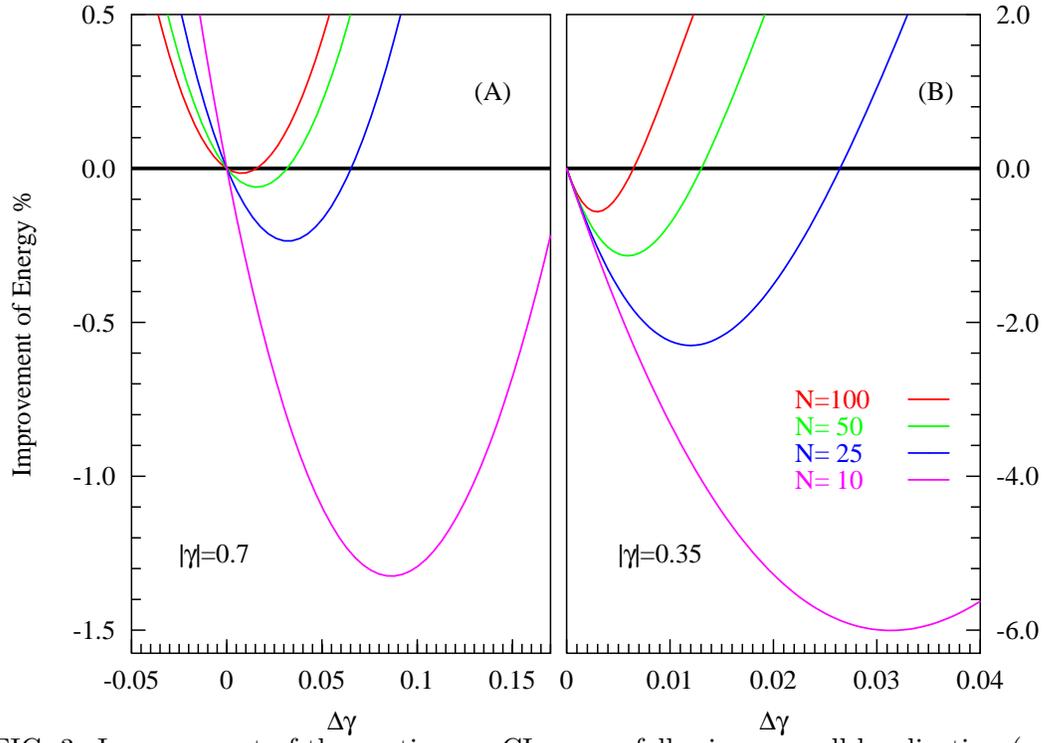}
\caption [kdv]{Improvement of the continuous CI energy following
a small localization (equivalent to positive values of $\Delta\gamma$) 
of the GP orbital for $N=10,25,50$ and $100$ bosons:
(A) $|\gamma|=0.7$;
(B) $|\gamma|=0.35$.
}
\end{figure}

\end{document}